\documentclass[12pt,a4paper
]{article}
\usepackage{latexsym}
\usepackage{amssymb}
\usepackage{graphicx}
\usepackage[cp1251]{inputenc}

\newcommand{\be}{\begin{equation}}
\newcommand{\ee}{\end{equation}}
\newcommand{\p}[1]{(\ref{#1})}
\def\a{\alpha}

\def\b{\beta}
\def\p{\partial}

\def\t{\theta}
\def\tr{{\rm tr}}
\def\tr{{\rm tr}\,}

\def\cN{{\cal N}}

\def\bea{\begin{eqnarray}}
\def\eea{\end{eqnarray}}

\def\cN{{\cal N}}

\def\f{\frac}
\def\n{\nabla}
\def\tr{{\rm tr}\,}

\def\bea{\begin{eqnarray}}
\def\eea{\end{eqnarray}}

\def\s{\sigma}

\def\d{\delta}
\def\q{\quad}
\def\g{\gamma}

\def\ve{\varepsilon}

\def\O{\Omega}

\date{\it  }
\begin{document}

\begin{center}
\vspace{1cm} {\Large\bf Towards harmonic superfield formulation of
$\cN=4$, $USp(4)$  SYM theory with the  central charge \vspace{1.2cm}

  }

\vspace{.2cm}
 { I.L. Buchbinder$^{a}$,  N.G. Pletnev$^{b}$

}

\vskip 0.6cm { \em \vskip 0.08cm \vskip 0.08cm $^{a}$Department of
Theoretical Physics, Tomsk State Pedagogical University,\\ Tomsk,
634061 Russia \vskip 0.08cm \vskip 0.08cm $^{b}$Department of
Theoretical Physics, Institute of Mathematics, \\ Novosibirsk,
630090 Russia

 }
\vspace{.2cm}
\end{center}

\begin{abstract}
We develop a superfield formulation of $\cN=4$ supersymmetric
Yang-Mills theory with the rigid central charge in $USp(4)$ harmonic
superspace. The component formulation of this theory was given by
Sohnius, Stelle and West \cite{SSW80}, but its superfield
formulation has not been constructed so far. We construct the
superfield action, corresponding to this model, and show that it
reproduces the component action from \cite{SSW80}.

\end{abstract}

\section{Introduction}

Maximally extended $\cN=4$ SYM theory with R-symmetry group $SU(4)$
possesses many remarkable properties on classical and quantum levels
and is widely explored in modern theoretical and mathematical
physics. For the first time this theory was obtained by dimensional
reduction of the ten dimensional $\cN=1$ SYM theory to the four
dimensions \cite{BSS}.  Field content of this theory involves one
vector, six real scalars and four Majorana spinors. By construction,
such model is non-manifestly supersymmetric and the supersymmetry
transformations are closed only on-shell. Its formulation in terms
of the on-shell superfields was given in \cite{sohnius78}. In many
cases, especially to study the quantum aspects, it would be
preferable to get an off-shell formulation of $\cN=4$ SYM theory. It
is generally accepted that off-shell formulation of the
supersymmetric theories is realized in terms of corresponding
unconstrained superfields (see e.g. \cite{1001}, \cite{Idea}).
However, in spite of the considerable efforts, superfield
formulation of $\cN=4$ SYM theory in terms of unconstrained $\cN=4$
superfields is still unknown. The best, that has been obtained so
far is its formulation in terms of $\cN=1$ superfields \cite{1001}
and $\cN=2,3$ harmonic superfields \cite{GIOS}, \cite{GIKOS}.
Attempts to develop a consistent unconstrained $\cN=4$ harmonic
superfield formulation for the $\cN=4$ SYM theory have not been
successful so far \cite{1}.

We would like to draw attention to another $\cN=4$ supersymmetric
model, namely $USp(4)$ SYM theory with central
charge\footnote{Structure of the extended supersymmetry theories
with the central charges is discussed in \cite{S78}.}. Due to the
central charge, the R-symmetry group of the corresponding
superalgebra is a subgroup $USp(4)$ of the group $SU(4)$. Field
content of such a gauge model involves one real vector, five real
scalars, four Majorana spinors and auxiliary fields - axial vector
and five real scalars. Furthermore, special constraints are imposed
on the auxiliary and dynamical fields. After eliminating the
auxiliary fields with the help of an additional scalar field, the
conventional $SU(4)$, $\cN=4$ SYM theory is restored. In Abelian
case the constraint is solved for the auxiliary vector field in
terms of an antisymmetric second rank field \cite{SSW80}. This field
describes a propagating spin-0 mode which is known as a 'notoph'
\cite{notoph}. As a result we get the conventional $\cN=4$ SYM
theory where one of the scalars is replaced by an antisymmetric
tensor field. In addition, with respect to central charge
transformations the vector field transforms into the dual field
strength of the antisymmetric tensor and the antisymmetric tensor
transforms into the dual field strength of the vector. Therefore
such a model can be treated as some kind of vector-tensor multiplet
theory\footnote{The $\cN=2$ vector-tensor multiplet theories are
discussed in \cite{deWit}, \cite{novak}, \cite{N2},
\cite{N2harmonic}.}.

In non-Abelian case this constraint is not solved in local form,
however the constraint can be inserted into action with help of a
scalar Lagrange multiplier\footnote{It was shown in \cite{3} that in
principle a local Lagrangian can be written but the resulting
Lorentz and supersymmetry transformations are non-local.}. It is
evident that such a procedure is not supersymmetric, the authors of
\cite{SSW80} expressed a hope that it can be made supersymmetric if
it becomes possible to construct a superfield formulation.

In this paper we develop a superspace formulation of $\cN=4$ SYM
theory with the rigid central charge \cite{SSW80} in terms of
$\cN=4$ superfields. Some superspace aspects of the theory under
consideration have been discussed in the earlier papers
\cite{Mil83}, \cite{Sait05}, \cite{BLS}. The paper \cite{Mil83} is
devoted to construction of the above theory in terms of $\cN=1$
superfields where the component constraint from \cite{SSW80} has
been written in $\cN=1$ superfield form. The paper \cite{Sait05}
develops a gauge theory in the $\cN=4$, $USp(4)$ superspace. Here
the superfield constraints, originating from the Bianchi identities
and determining the correct $USp(4)$ vector multiplet with the
central charge have been formulated and solved in terms of some
superfield strengths. In the papers \cite{BLS} the aspects of a
$\cN=4$ SYM theory in harmonic superspace with the central charge
were studied and applied for the construction of Abelian low-energy
effective action. In the present paper we prove that the constraint
for the auxiliary field, introduced in \cite{SSW80} for non-Abelian
theory, automatically follows from the $\cN=4$ superfield
constraints stipulated by the Bianchi identities, obtained in
\cite{Sait05}. Also, we develop a $USp(4)$, $\cN=4$ harmonic
superspace formalism and propose a gauge invariant, $\cN=4$
supersymmetric action, which exactly reproduces the component action
of \cite{SSW80} for non-Abelian theory.

The paper is organized as follows. In subsection 1.1 the theory of
Sohnius, Stelle and West is briefly discussed. Section 2 is devoted
to the structure of the constraints and their solutions in
conventional $USp(4)$, $\cN=4$ superspace. Here we show how the
component constraint from \cite{SSW80} is derived from the
constraints on superfield strengths. In section 3 we reformulate
these constraints in $USp(4)$, $\cN=4$ harmonic superspace, define
the corresponding analytic subspace, propose the $\cN=4$
supersymmetric action and find its component form. Section 4 briefly
summarizes the results. The notations, conventions and some details
of computations are given in Appendices A, B and C.

\subsection{$USp(4)$ SYM model with the central charge}
$\cN$=4, $USp(4)$ SYM model with the central charge has been
proposed by Sohnius, Stelle and West in the component formulation
\cite{SSW80}. This model possesses  the $USp(4)$ R-symmetry and is
described by the action\footnote{Here we follow the notations from
\cite{SSW80}. Matrix $\Omega$, associated with $USp(4)$ symmetry, is
given in Appendix A.} \be\label{act comp} S=\tr\int d^4x (-\f14
F_{mn}F^{mn}-\f12V_mV^m+\f12
\n_m\phi_{ij}\n^m\phi^{ij}+\f12H_{ij}H^{ij} \ee
$$-\f{i}{4}\bar\lambda^i\not\!\n\lambda_i-\bar\lambda^i[\lambda^j,\phi_{ij}]+
\f14[\phi_{ij},\phi_{kl}][\phi^{ij},\phi^{kl}])~.$$ Here the vector
field $A_m$, $USp(4)$-Majorana spinor fields
$\lambda_{i\a}=\bar\lambda^{j\b}(C^{-1})_{\b\a}\O_{ji}$ and the
antisymmetric, $\Omega$-traceless scalar fields $\phi_{ij}$ are the
propagating fields whereas the pseudovector $V_m$ and antisymmetric,
$\Omega$-traceless, scalar fields $H_{ij}$ are the auxiliary fields.
All fields take the values in the Lie algebra of a gauge group.
$\n_m$ are the conventional gauge covariant derivatives.

Supersymmetry transformations of the component fields are defined as
follows
\be\label{supertrans}\d
A_m=i\bar\epsilon^i\g_m\lambda_i, \q
\d\phi_{ij}=-i(\bar\epsilon_{[i}\lambda_{j]}+\f12\Omega_{\ij}\bar\epsilon^k\lambda_k)~,\ee
$$\d\lambda_i=-\f12\s_{mn}F^{mn}\epsilon_i+2\not\!\n\phi_i^{\ \ j}\epsilon_j
+\g^5\g^mV_m\epsilon_i+2\g^5H_i^{\ \ j}\epsilon_j-2i[\phi_{ik},\phi^{kj}]\epsilon_j~,$$
$$\d H_{ij}=i(\bar\epsilon_{[i}\g^5\not\!\n\lambda_{j]}+
\f12\Omega_{ij}\bar\epsilon^k\g^5\not\!\n\lambda_k)+\bar\epsilon^k\g^5[\lambda_k,\phi_{ij}]$$
$$-2(\bar\epsilon_{[i}\g^5[\lambda^k,\phi_{j]k}]+\f12\Omega_{ij}\bar\epsilon^{k}\g^5[\lambda^l,\phi_{kl}])~,$$
$$\d V_m=i\bar\epsilon^i\g^5\s_{mn}\n^n\lambda_i+2\bar\epsilon^i\g^5\g_m[\lambda^k,\phi_{ik}]~.$$
The supersymmetry transformations (\ref{supertrans}) are closed
off-shell, up to field dependent gauge transformations. The
anticommutator of two supersymmetry transformations leads to
coordinate translations, central charge transformations and gauge
transformations on the gauge vector fields and the physical scalar
fields. The transformations generated by the central charge are
defined in the form \be\label{centrtrans}\d_zA_m=\omega V_m,\q
\d_z\phi_{ij}=-\omega H_{ij}~,\ee
$$\d_z\lambda_i=-\omega(\g^5\not\!\n\lambda_i-2i\g^5[\lambda^k,\phi_{ik}])~,$$
$$\d_zV_m=\omega(\n^nF_{nm}-\f12\{\bar\lambda^k,\g_m\lambda_k\}+i[\phi^{kl},\n_m\phi_{kl}])~,$$
$$\d_zH_{ij}=\omega(-\n^m\n_m\phi_{ij}+\f{i}{4}\Omega_{ij}\{\bar\lambda_k,\lambda^k\}
-[\phi_{kl}, [\phi^{kl},\phi_{ij}]])~.$$

The action (\ref{act comp}) for the $USp(4)$ multiplet  is invariant
under the supersymmetry transformations if, and only if, the
following additional non-linear constraint
\be\label{costrV}\n^mV_m+\f12\{\bar\lambda^i,\g^5\lambda_i\}-i[\phi_{ij},
H^{ij}]=0~,\ee is satisfied. Moreover, the action (\ref{act comp})
is invariant under central charge transformations (\ref{centrtrans})
with the constant  parameter $\omega.$

The conventional on-shell $SU(4)$ ${\cal N} =4$ SYM theory \cite{BSS}, \cite{sohnius78} is
obtained by introducing the scalar Lagrange multiplier $A_5$ for the
constraint and eliminating the auxiliary fields $V_m$ and $H_{ij}$
from the equations of motion
\be\label{Legandr}V_m=-\n_mA_5,\q
H_{ij}=i[A_5,\phi_{ij}]~.
\ee
In this case, the scalar field $A_5$ is
unified with 5 scalar fields ${\phi}_{ij}$ and as a result one gets
6 scalar fields of conventional $SU(4)$ ${\cal N} =4$ SYM theory.

The aim of the paper is to develop a formulation of the model under
consideration in terms of $\cN=4$ harmonic superfields and present
the action in superfield form.

\section {$USp(4)$, $\cN=4$  superspace and SYM model with  central charge}
The $\cN$=4 central charge superspace is represented by the
coordinates $Z^M=\{x^m, z, \t^\a_i, \bar\t^i_{\dot\a}\}$ and the
supercovariant derivatives $D_M=(\p_m,\p_z, D^i_\a,
\bar{D}^{\dot\a}_i)$ \footnote{We mainly use the notations and
conventions from \cite{Idea} and \cite{GIOS}. See Appendix A for
details.}. These derivatives are used for the definitions of the
gauge covariant derivatives $\n_M=D_M+i\Gamma_M$ with
superconnections $\Gamma_M$ and gauge transformations
$\n'_M=e^{i{\tau}}\n_M e^{-i{\tau}},$ where ${\tau}$ is a gauge
superfield parameter. Then one introduces the curvature tensors or
superfield strengths defined on the $USp(4)$ ${\cal N}=4$ central
charge superspace with the help of algebra:
\be\label{supalg}\{\n_{\a i},\n_{\b
j}\}=2i\ve_{\a\b}\O_{ij}\n_z+2i\ve_{\a\b}W_{ij}, \q \{\bar\n_{\dot\a
i},\bar\n_{\dot\b
j}\}=2i\ve_{\dot\a\dot\b}\O_{ij}\n_{{z}}-2i\ve_{\dot\a\dot\b}W_{ij}~,\ee
$$\{\n_{\a i}, \bar\n_{\dot\a j}\}=-2i\O_{ij}\n_{\a\dot\a}~,
\footnote{These gauge supercovariant derivatives obey the usual
constraints
$$\{\n^{(i}_{\hat{\a}}, \n^{j)}_{\hat{\b}}\}=0,
\q \{\n_\a^i, \bar\n_{\dot\a j}\}-\f14\delta^i_j\{\n_\a^k, \bar\n_{\dot\a k}\}=0~.
$$ Here and below we use the notation $\hat{\a}$ for the set $\{\a, \dot\a\}~.$
}$$
where we impose the reality conditions under the internal
symmetry:
\be\label{real}\Omega^{ij}W_{ij}=0,\q
\overline{(W_{ij})}=W^{ij}=\O^{ik}\O^{jl}W_{kl}=-\f12\ve^{ijkl}W_{kl}~.\ee
Here $\O_{ij}$ is the invariant tensor of the $USp(4)$ group. The
other commutators of the gauge supercovariant derivatives look like
\be\label{constraints}
[\n_{\a i}, \n_z]=iG_{\a i},\q [\bar\n_{\dot\a
i}, \n_z]=-i\bar{G}_{\dot\a i}~,\ee
$$[\n_{\a i}, \n_m]=iF_{\a i m},\q [\bar\n_{\dot\a i},
\n_m]=-i\bar{F}_{\dot\a i m}~,$$
$$ [\n_m,\n_z]=iV_m, \q
[\n_m,\n_n]=iF_{mn}~.$$ In this representation $G_{\a i}$ is the
$USp(4)$  Majorana spinor with the reality condition
$\overline{(G_{\a i})}=\bar{G}_{\dot\a}^i=\O^{ij}\bar{G}_{\dot\a
i}$. As a result the gauge theory in $USp(4)$, ${\cal N}=4$ central
charge superspace is described by the superfields  $W_{ij},
G_{\hat\a i}, F_{\hat\a i m},  V_m, F_{mn}$\footnote{Do not confuse
the superfields $V_m, F_{mn}$ and others in relations
(\ref{constraints}) with fields $V_m, F_{mn}$ in the component
action (\ref{act comp}). Actually component fields $\phi_{ij},
\lambda^i_{\hat\a},  V_m, F_{mn}$ will be the lowest components of
the superfields $W_{ij}, G^i_{\hat\a}, V_m, F_{mn}$.}. The
superfield strengths satisfy some number of the constraints to
reduce the number of fields to an irreducible multiplet stipulated
by the Bianchi identities. The solution of these relations
determines the field content of the theory as well as the
transformation laws of the component fields.

\subsection{Solving the Constraints}
One can prove that the Bianchi identities are satisfied if, and only
if, all superfield strengths are expressed in terms of a single real
scalar superfield $W_{ij}$ and its spinor derivatives. Here we only
list the results of \cite{Sait05} in our conventions concerning the
solution to the constraints of the dimension from 3/2 to 3:

$\bullet$ \ \ \ Solution to the $dim=\f32$ Bianchi identities:

\be\label{F-G} F_{\a i m}=-\s^m_{\a\dot\a}\bar{G}^{\dot\a}_i, \q\q
\bar{F}_{\dot\a i m}=G^\a_i\s^m_{\a\dot\a}~,\ee
\be\label{defG}\n_{\a k}W_{ij}=i\O_{ij}G_{\a k}+2i\O_{k[i}G_{\a j]},
\q \bar\n_{\dot\a k}W_{ij}=i\O_{ij}\bar{G}_{\dot\a
k}+2i\O_{k[i}\bar{G}_{\dot\a j]}~,\ee
\be\label{H} 5i
G_{\hat{\a}i}=\n^k_{\hat{\a}}W_{ki}, \q\q \n_zW_{ij}\equiv
H_{ij}~.\ee

$\bullet$ \ \ \ Solution to the $dim=2$ Bianchi identities:

\be\label{consG}\n_{\a i}G_{\b
j}=-\ve_{\a\b}H_{ij}-\f12\O_{ij}F_{\a\b}+\f12\ve_{\a\b}[W_{ik},
W_{j}^{\ \ k}]~, \footnote{Note that in the Abelian case the
superfields $G_\a^i, \bar{G}^i_{\dot\a}$ satisfy the constraints
$D_\a^{(i}G_\b^{j)}=\bar{D}_{\dot\a}^{(i}G_\b^{j)}=0$.}\ee
$$\bar\n_{\dot\a i}\bar{G}_{\dot\b
j}=-\ve_{\dot\a\dot\b}H_{ij}+\f12\O_{ij}\bar{F}_{\dot\a\dot\b}-\f12\ve_{\dot\a\dot\b}[W_{ik},
W_j^{\ \ k}]~,$$
\be\label{eqG}\bar\n_{\dot\a i}G_{\a
j}=i\O_{ij}V_{\a\dot\a}-\n_{\a\dot\a}W_{ij}, \q\q \n_{\a
i}\bar{G}_{\dot\a j}=i\O_{ij}V_{\a\dot\a}+\n_{\a\dot\a}W_{ij}~.\ee

$\bullet$ \ \ \ Solution to the $dim=\f52$ Bianchi identities:

\be\label{Dir}\n_z\bar{G}_{\dot\a i}=\n_{\a\dot\a}G^\a_i+[W_{ik},
\bar{G}^k_{\dot\a}],\q\q \n_z G^\a_i=\n^{\dot\a\a}\bar{G}_{\dot\a
i}-[W_{ik}, G^{\a k}]~,\ee
\be\n_{\a i}V_m=\s^m_{\a\dot\a}[W_{ik},\bar{G}^{\dot\a
k}]+i(\s_{mn})_\a^\b\n_n G_{\b i}~,\ee
$$ \bar\n_{\dot\a
i}V_m=-\s^m_{\a\dot\a}[W_{ik}, G^{\a k}]+i\n_n\bar{G}_{\dot\b
i}(\bar\s_{mn})^{\dot\b}_{\dot\a}~,$$
$$\n_{\a i}H_{jk}=-i\O_{jk}\n_{\a\dot\a}\bar{G}^{\dot\a}_i-2i\O_{i[j}\n_{\a\dot\a}\bar{G}^{\dot\a}_{k]}-i[W_{jk},G_{\a i}]
-i\O_{jk}[W_{il},G^l_\a]-2i\O_{i[j}[W_{k]l},G^l_\a]~,$$
$$\bar\n_{\dot\a i}H_{jk}=i\O_{jk}\n_{\a\dot\a}G^\a_i+2i\O_{i[j}\n_{\a\dot\a}G^\a_{k]}+i[W_{jk},\bar{G}_{\dot\a i}]
+i\O_{jk}[W_{ip}, \bar{G}^p_{\dot\a}]+2i\O_{i[j}[W_{k]p}, \bar{G}^p_{\dot\a}]~,$$
\be\n_{\dot\a i}F_{\a\b}=2i\n_{(\a\dot\a}G_{\b) i},  \q\q \n_{\g i}F_{\a\b}=-2i\ve_{\g(\a}\n_{\b)\dot\a}\bar{G}^{\dot\a}_i~,\ee
$$\n_{\a i}F_{\dot\a\dot\b}=2i\n_{\a(\dot\a}\bar{G}_{\dot\b) i}, \q\q \bar\n_{\dot\g i}F_{\dot\a\dot\b}=
-2i\ve_{\dot\g(\dot\a}\n_{\a\dot\b)}{G}^{\a}_i~.$$

$\bullet$  \ \ \ Solution to the $dim=3$ Bianchi identities:

\be\label{DV}\n_mV_m=-\f{i}{4}[W_{ik},H^{ik}]+\f12\{G_{\a k},G^{\a
k}\}+\f12\{\bar{G}_{\dot\a k},\bar{G}^{\dot\a k}\}~, \ee
\be\label{eomF}\n_zV_m=-\s^m_{\a\dot\a}\{G^{\a
i},\bar{G}_i^{\dot\a}\}-\f{i}{4}[W_{ik},\n_mW^{ik}]-\n_aF_{am}~,\ee
\be\label{eomW} \n_zH_{jk}=\Box W_{jk}-\f{i}{2}\O_{jk}\{G_{\a
i},G^{\a i}\}-2i\{G_{\a j},G^\a_k\}+\f{i}{2}\O_{jk}\{\bar{G}_{\dot\a
i},\bar{G}^{\dot\a i}\}+2i\{\bar{G}_{\dot\a j},
\bar{G}^{\dot\a}_k\}\ee
$$+\f{1}{8}\O_{jk}[W_{il}[W^i_{\ \ p},W^{lp}]]~.$$

Using the covariant derivatives and solving the Bianchi identities,
we can immediately write down the supersymmetry transformations
(\ref{supertrans}) of the component fields in the form $\d \Phi
|=-\epsilon^{\hat\a i}\n_{\hat\a i} \Phi |,$ where $\Phi$ is any of
the superfields under consideration. The central charge
transformations (\ref{centrtrans}) are realized on $\n_M$ and
superfields $\Phi$ as $\d_z \n_M=[\omega\n_z,\n_M] $ and
$\d_z\Phi=\omega \n_z\Phi$ with $\omega$ being a constant parameter,
corresponding to global transformation of  the central charge
coordinate $\d z=\omega$.

By analogy with the case of the $\cN = 2$ models with the  intrinsic
central charge \cite{deWit}-\cite{N2harmonic}, the equations
(\ref{Dir}, \ref{eomF}, \ref{eomW}) can be called the generalized
Dirac equation, the Yang-Mills equation  and the Klein-Gordon
equation respectively. Thus the central charge plays the role of
'fifth coordinate'. As a consequence of (\ref{F-G})-(\ref{eomW}), we
can construct the complete power expansion of the
$W^{ij}=\sum_{k=0}^\infty W^{ij}_{(k)} z^k$ in superspace from the
first two coefficients $W^{ij}_{(0)}, W^{ij}_{(1)}$. The set of
constraints (\ref{F-G})-(\ref{eomW}) shows that the component fields
in \cite{SSW80} are completely determined by the lowest orders in
the expansion $W_{ij}(x,\t,\bar\t,z)$ over $z, \t, \bar\t$.  Then it
is clear that we can completely fix the dependence of all quantities
under consideration on the central-charge coordinate $z$, as was
done for the theories with $\cN=2$ rigid supersymmetry with the
central charge \cite{deWit}, \cite{N2}, \cite{N2harmonic}.

Now we clarify how the conditions (\ref{Legandr}) appear in the
superfield approach. It is known from \cite{SSW80} that the
equations (\ref{Legandr}) allow us to reduce the ${\cal N}=4,
USp(4)$ SYM theory to the conventional ${\cal N}=4, SU(4)$ SYM
theory. The relations (\ref{Legandr}) are the solutions of equations
of motion for auxiliary fields $V_m, H_{ij}$ with help of constraint
(\ref{costrV}). Therefore, if we want to get the analogous relation
between the above two theories in the superfield approach, we also
should use, besides the identities (\ref{F-G}) - (\ref{DV}), some
additional equations or restrictions. We take such restrictions in
the form \be\label{restrictions}
\partial_{z}\Gamma_{m,z}=0, \q \p_zW_{ij}=0 ~\ee
and show that these conditions are equivalent to (\ref{Legandr}).
Consider the superfields $V_m$ and $H_{ij}$ defined by the
identities (\ref{constraints}) and (\ref{H}): $iV_m=[\n_m,\n_z],
H_{ij}=[\n_z,W_{ij}]$. Let $V_m\sim V_m|_{\t=0}$, $\phi_{ij} \sim
W_{ij}|_{\t=0}$, $A_{5} \sim \Gamma_{z}|_{\t=0}$ and use the
constraints (\ref{restrictions}). Then we immediately get the
relations (\ref{Legandr}). As a result, we see how the conditions
(\ref{Legandr}) are obtained in the superfield approach. They are
the consequences of the restrictions (\ref{restrictions}), which
should be added to the identities (\ref{constraints}), (\ref{H}) if
we want to reduce the $ USp(4)$ SYM theory to the on-shell $SU(4)$
SYM theory. In this case the identities (\ref{Dir}), (\ref{DV}),
(\ref{eomF}), (\ref{eomW}) are converted into equations of motion
for conventional ${\cal N}=4$ SYM theory. If we do not impose the
conditions (\ref{restrictions}), we have the $ USp(4)$ superfield
gauge theory satisfying  the identities (\ref{F-G}) -- (\ref{eomW}).

The constraints  (\ref{defG}) allow us to derive the important
consequences. Specifying concrete values of the indices $i,j,k$  and
using explicit form of the matrix $\Omega$ one obtains the various
constraints for the superfield $W_{ij}$. For example
\be\label{constW} \n_{\a 1}W_{12}=0, \q \n_{\a 2}W_{12}=0~,\ee and
the other analogous equations. These equations mean the very special
dependence of the superfield $W_{ij}$ on anticommuting coordinates
\be W_{12}=W_{12}(\t^3, \t^4, \bar\t_2, \bar\t_1),\q
W_{13}=W_{13}(\t^2,\t^4,\bar\t_3,\bar\t_1)~, \ee
$$W_{24}=W_{24}(\t^1,\t^3,\bar\t_4,\bar\t_2),\q W_{34}=W_{34}(\t^1,\t^2, \bar\t_4,\bar\t_3)~,$$
$$W_{14}(\t^i,\bar\t_i)+W_{23}(\t^i,\bar\t_i)=0~.$$
The coordinates $x^{m}$  and $z$ are not written down. As a result
we see that $W_{ij}$ with different values $i,j$ belong to different
subspaces of the full superspace. A more transparent covariant
solution of such constraints will take place in the framework of the
harmonic superspace.

\section{Gauge theory in $USp(4)$
harmonic superspace}
\subsection{Harmonic formalism}
Following the general scheme of harmonic superspace construction
\cite{GIOS}\footnote{Structure of the harmonic variables for
$\cN=2,3,4$ harmonic superspaces with various R-symmetries is
discussed in \cite{IKVO85}.} we extend the $\cN=4$ central charge
superspace with coordinates $Z^M=(x^m,z, \t_{\a
i},\bar\t^i_{\dot\a})$, by the eight-dimensional coset space
$USp(4)/U(1)\times U(1)$ parametrized by the harmonic variables
$u_i^{(\pm,0)}, u_i^{(0,\pm)},$ which are inert under supersymmetry
and take the values in the fundamental representation of $USp(4)$
\cite{IKVO85}, \cite{BLS} (see Appendix B for details) \footnote{
Alternative formalism for the superfield description of this theory
can, in principle, be based on the $SO(5)/U(2)$ harmonic superspace
or on the $SO(5)/U(1)\times U(1)$ harmonic superspace \cite{so5}.}.

Using the harmonics  $u_i^{(\pm,0)}, u_i^{(0,\pm)},$ we define the
harmonic derivatives $\partial^{(q_1,q_2)}$, which are left-invariant vector fields on $USp(4)$, by the rule
\be\label{Hpart} \p^{(\pm\pm,0)}=u_i^{(\pm,0)}\f{\p}{\p u_i^{(\mp,0)}}, \q\q \p^{(0,\pm\pm)}=
u_i^{(0,\pm)}\f{\p}{\p u_i^{(0,\mp)}}~,\ee
$$ \p^{(\pm,\pm)}=u_i^{\pm,0}\f{\p}{\p u_i^{0,\mp}}+u_i^{0,\pm}\f{\p}{\p u_i^{\mp,0}},\q \q \p^{(\pm,\mp)}=
u^{(\pm,0)}_i\f{\p}{\p u_i^{0,\pm}}-u^{(0,\mp)}_i\f{\p}{\p
u_i^{\mp,0}}~.$$ The commutation relations for the step operators
(\ref{Hpart}) together with the two Cartan generators of the algebra
$usp(4)$  are given by (\ref{harmalg}). The algebra $usp(4)$ is
ten-dimensional and its rank equals two. Therefore there are 4
positive roots, 2 of which are simple roots and similarly for the
negative roots. Also, there is a special involution (special complex
conjugation)
\be\label{tilde}\widetilde{u_i^{(\pm,0)}}=u^{(0,\pm)i},\q
\widetilde{u^{(0,\pm)i}}=-u_i^{(\pm,0)},\ee and so on, allowing us
to define a reality condition in harmonic superspace \cite{IKVO85},
\cite{BLS}.

With the help of the harmonics $u_i^{(\pm,0)},\q u_i^{(0,\pm)}$ one
can convert the spinor covariant derivatives into the operators
\be\n^{(\pm,0)}_{\hat{\a}}=u^{(\pm,0)}_i\n^{i}_{\hat{\a}},\q
\n^{(0,\pm)}_{\hat{\a}}=u_i^{(0,\pm)}\n^{i}_{\hat{\a}}~,\ee and
reformulate the superalgebra (\ref{supalg}) in another, clearer
form. Now we can rewrite the constraints (\ref{constW}) as the
constraints in harmonic superspace where they will have a form of
analyticity conditions.

In the $\tau$-frame we have  the obvious  anticommuting
relations
\be\label{GA}\{\n_{\hat{\a}}^{(+,0)},\n_{\hat{\b}}^{(+,0)}\}=0,
\q \{\n_{\hat{\a}}^{(0,+)},\n_{\hat{\b}}^{(0,+)}\}=0~,\ee
$$\{\n_\a^{(\pm,0)}, \bar\n_{\dot\a}^{(\mp,0)}\}=\mp 2i\n_{\a\dot\a},\q
\{\n_\a^{(0,\pm)},
\bar\n_{\dot\a}^{(0,\mp)}\}=\mp2i\n_{\a\dot\a}~,$$ and the other
anticommuting relations that define five harmonic projection of the
tensor $W^{ij}$\footnote{The upper sign here corresponds to $\alpha$
in $\hat{\a}$ and the lower sign corresponds to $\dot{\a}$ in
$\hat{\a}$.}
\be\label{defW}\{\n^{(+,0)}_{\hat{\a}},\n^{(-,0)}_{\hat{\b}}\}=2i\ve_{\hat{\a}\hat{\b}}\n_z\pm
2i\ve_{\hat{\a}\hat{\b}}W_1^{(0,0)},\q
 \{\n^{(0,+)}_{\hat{\a}},\n^{(0,-)}_{\hat{\b}}\}=
 2i\ve_{\hat{\a}\hat{\b}}\n_z\pm 2i\ve_{\hat{\a}\hat{\b}}W_2^{(0,0)}~,\ee
$$ \{\n^{(+,0)}_{\hat{\a}},\n^{(0,-)}_{\hat{\b}}\}=\pm 2i\ve_{\hat{\a}\hat{\b}}W^{(+,-)},\q
\{\n^{(+,0)}_{\hat{\a}},\n^{(0,+)}_{\hat{\b}}\}=\pm
2i\ve_{\hat{\a}\hat{\b}}W^{(+,+)}~,
$$
$$ \{\n^{(-,0)}_{\hat{\a}},\n^{(0,+)}_{\hat{\b}}\}=\pm 2i\ve_{\hat{\a}\hat{\b}}W^{(-,+)},
\q \{\n^{(-,0)}_{\hat{\a}},\n^{(0,-)}_{\hat{\b}}\}=\pm
2i\ve_{\hat{\a}\hat{\b}}W^{(-,-)}~,$$
where:
\be
u^{(+,0)}_iu^{(-,0)}_jW^{ij}=W_1^{(0,0)},\q
u^{(+,0)}_iu^{(0,+)}_jW^{ij}=W^{(+,+)},\q
u^{(+,0)}_iu^{(0,-)}_jW^{ij}=W^{(+,-)}~,\ee
$$ u^{(-,0)}_iu^{(0,+)}_jW^{ij}=W^{(-,+)},\q u^{(-,0)}_iu^{(0,-)}_jW^{ij}=
W^{(-,-)},\q u^{(0,+)}_iu^{(0,-)}_jW^{ij}=W^{(0,0)}_2=-W^{(0,0)}_1~.$$
These definitions simply mean that $W^{(q_1,q_2)}$ in $\tau$-basis
depend linearly on harmonics $u^{(\pm,0)}_i,u^{(0,\pm)}_i$ and all
five harmonic projections of $W_{ij}$ transform through each other
under the action of symmetry generators. In particular
\be\label{hshort}\p^{(++,0)}W^{(+,+)}=0, \q \p^{(++,0)}W^{(-,+)}=W^{(+,+)}, \q \p^{(+,-)}W^{(-,+)}=-2W_1^{(0,0)}~, \ee
$$\p^{(--,0)}W^{(+,+)}=W^{(-,+)},\q \p^{(0,--)}W^{(+,+)}=W^{(+,-)},\q \p^{(-,-)}W^{(+,+)}=2W_1^{(0,0)}~, $$
$$\p^{(+,-)}W^{(+,+)}=\p^{(-,+)}W^{(+,+)}=0~.$$

The $ USp(4)$, $\cN=4$ harmonic superspace with coordinates $\{x^m,
z, \t^{(\pm,0)}_{\hat{\a}}, \t^{(0,\pm),}_{\hat{\a}},u_i\}$ contains
several analytic subspaces of the full superspace with eight
anticommuting coordinates. It can be checked that each of following
four superfields lives in its own analytic subspace

$\bullet$ $$ W^{(+,+)}
(\t^{(+,0)},\t^{(0,+)},\bar\t^{(+,0)},\bar\t^{(0,+)})~,$$

\be\label{harmonic constraints}\n_\a^{(+,0)}W^{(+,+)}=\n_\a^{(0,+)}W^{(+,+)}=\bar\n_{\dot\a}^{(+,0)}W^{(+,+)}=\bar\n_{\dot\a}^{(0,+)}W^{(+,+)}=0~,
\ee

$\bullet$
$$W^{(+,-)}(\t^{(+,0)},\t^{(0,-)},
\bar\t^{(+,0)},\bar\t^{(0,-)})~,$$
\be\label{harmonic constraints1}\n^{(+,0)}_\a W^{(+,-)}=\n^{(0,-)}_\a W^{(+,-)}=\bar\n^{(+,0)}_{\dot\a} W^{(+,-)}=
\bar\n^{(0,-)}_{\dot\a} W^{(+,-)}=0~,\ee

$\bullet$
$$W^{(-,+)}(\t^{(-,0)}, \t^{(0,+)},
\bar\t^{(-,0)}, \bar\t^{(0,+)})~,$$
\be\label{harmonic constraints2}\n^{(-,0)}_\a W^{(-,+)}=\n^{(0,+)}_\a W^{(-,+)}=\bar\n^{(-,0)}_{\dot\a} W^{(-,+)}=
\bar\n^{(0,+)}_{\dot\a} W^{(-,+)}=0~,\ee

$\bullet$ $$W^{(-,-)}(\t^{(-,0)}, \t^{(0,-)}, \bar\t^{(-,0)},
\bar\t^{(0,-)})~,$$ \be\label{harmonic constraints3}\n^{(-,0)}_\a
W^{(-,-)}=\n^{(0,-)}_\a W^{(-,-)}=\bar\n^{(-,0)}_{\dot\a} W^{(-,-)}=
\bar\n^{(0,-)}_{\dot\a} W^{(-,-)}=0~.\ee The coordinates $x^m, z, u$
are manifestly not written down here. The relations (\ref{harmonic
constraints})-(\ref{harmonic constraints3}) form the full list of
constraints stipulated by the relations (\ref{defG}). In each of the
above analytic subspaces the corresponding spinor derivatives become
short. However, the dependence of the superfield
$W^{(0,0)}(\t^{(\pm,0)}, \t^{(0,\pm)}, \bar\t^{(\pm,0)},
\bar\t^{(0,\pm)})$ on anticommuting coordinates is not restricted.

Acting on $W^{(q_1,q_2)}$ by one, two, three or four spinor
derivatives $D^{(q_1,q_2)}_{\hat{\a}}$, one obtains a set of
relations which allow one to define the superstrength components. We
do not write down all such relations. The relations that will be of
further use have the form \be\label{harmBianch}
G^{(+,0)}_{\hat{\a}}=i\n^{(+,0)}_{\hat{\a}}
W^{(0,0)}_1=\f{i}{2}\n^{(0,+)}_{\hat{\a}}
W^{(+,-)}=-\f{i}{2}\n^{(0,-)}_{\hat{\a}} W^{(+,+)}~, \ee
$$G^{(0,+)}_{\hat{\a}}=-i\n^{(0,+)}_{\hat{\a}} W^{(0,0)}_1=
-\f{i}{2}\n^{(+,0)}_{\hat{\a}} W^{(-,+)}=\f{i}{2} \n^{(-,0)}_{\hat{\a}} W^{(+,+)}~,$$
\be\label{harmBianch1}\n_{\hat{\a}}^{(+,0)}G^{(+,0)}_{\hat{\b}}=\mp\ve_{\hat{\a}\hat{\b}}[W^{(+,+)},W^{(+,-)}]~,\ee
$$\n^{(-,0)}_{\hat{\a}}G^{(+,0)}_{\hat{\b}}=
\ve_{\hat{\a}\hat{\b}}H^{(0,0)}\pm\f12F_{\hat{\a}\hat{\b}}
\mp\f12\ve_{\hat{\a}\hat{\b}}\{[W^{(-,+)},W^{(+,-)}]+[W^{(+,+)},W^{(-,-)}]\}~,$$
$$\n^{(0,-)}_{\hat{\a}}G^{(+,0)}_{\hat{\b}}=
\ve_{\hat{\a}\hat{\b}}H^{(+,-)}\mp\ve_{\hat{\a}\hat{\b}}[W^{(0,0)},W^{(+,-)}],
\q \bar\n^{(0,-)}_{\dot\a}G^{(+,0)}_\a=\n_{\a\dot\a}W^{(+,-)}~,$$
$$\n^{(-,0)}_{\hat{\a}} G^{(0,+)}_{\hat{\b}}=
-\ve_{\hat{\a}\hat{\b}}H^{(-,+)}\mp\ve_{\hat{\a}\hat{\b}}[W^{(0,0)},W^{(-,+)}]~,$$
$$\n^{(-,0)}_\a\bar{G}^{(+,0)}_{\dot\a}=-iV_{\a\dot\a}-\n_{\a\dot\a}W^{(0,0)},\q \bar\n^{(-,0)}_{\dot\a}{G}^{(+,0)}_{\a}=
-iV_{\a\dot\a}+\n_{\a\dot\a}W^{(0,0)}~,$$
$$\n_\a^{(-,0)}V_m=\s^m_{\a\dot\a}\{[W^{(0,0)},\bar{G}^{(-,0)\dot\a}]-
[W^{(-,+)},\bar{G}^{(0,-)\dot\a}]+[W^{(-,-)},\bar{G}^{(0,+)\dot\a}]\}+i(\s_{mn})_\a^\b\n_nG^{(-,0)}_\b~.$$
Proceeding this way, we can find, in principle, the components of
all harmonic projections of the superfield $W_{ij}$ \footnote{It is
easy to prove, analogously to \cite{short}, that $W^{(+,+)}$  in the
absence of the central charge also satisfies the conditions of
linearity with respect to each variable $\t$:
$D^{(-,0)\hat{\a}}D^{(-,0)}_{\hat{\a}} W^{(+,+)}=0,$ since
$D^{(++,0)}D^{(-,0)\a}D^{(-,0)}_\a W^{(+,+)}=D^{(-,0)\a}D^{(+,0)}_\a
W^{(+,+)}=0$. }.

Further, we take the superfield $W^{(+,+)}$ as the basic superfield
strength and construct superfield action in its terms. This
superfield is a function on the analytic subspace of the harmonic
superspace parameterized by \be\label{anbasis}\{\zeta^M,
u\}=\{x^m_A,z_A,\t^{(+,0)}_\a, \t^{(0,+)}_\a,
\bar\t^{(+,0)}_{\dot\a}, \bar\t^{(0,+)}_{\dot\a}, u_i^{(\pm,0)},
u_i^{(0,\pm)}\}~,\ee where \be
x_A^m=x^m-i\t^{(-,0)}\s^m\bar\t^{(+,0)}-i\t^{(+,0)}\s^m\bar\t^{(-,0)}-i\t^{(0,-)}\s^m\bar\t^{(0,+)}-
i\t^{(0,+)}\s^m\bar\t^{(0,-)}~,\ee
$$ z_A=z+i\t^{(-,0)\a}\t^{(+,0)}_\a+i\t^{(0,-)\a}\t^{(0,+)}_\a-i\bar\t^{(+,0)}_{\dot\a}\bar\t^{(-,0)\dot\a}-
i\bar\t^{(0,+)}_{\dot\a}\bar\t^{(0,-)\dot\a}~.$$ One can prove that
the above analytic subspace is closed under the supersymmetry
transformations and is real with respect to 'tilde'-conjugation
(\ref{tilde}). Hence, the analytic superfield $W^{(+,+)}$ also can
be chosen real.

In the $\lambda$-frame the covariant spinor derivatives are
\be
D^{(+,0)}_{\hat{\a}}=\f{\p}{\p\t^{(-,0)\hat{\a}}}, \q
D^{(0,+)}_{\hat{\a}}=\f{\p}{\p\t^{(0,-)\hat{\a}}}~, \ee
$$ D^{(-,0)}_{{\a}}=-\f{\p}{\p \t^{(+,0){\a}}}+2i\bar\t^{(-,0)\dot\a}\p^A_{\a\dot\a}
-2i\t^{(-,0)}_\a\p^A_z,\q D^{(0,-)}_\a=-\f{\p}{\p \t^{(0,+)\a}}+
2i\bar\t^{(0,-)\dot\a}\p^A_{\a\dot\a}-2i\t^{(0,-)}_\a\p^A_z~,$$
$$\bar{D}^{(-,0)}_{\dot\a}=-\f{\p}{\p \bar\t^{\dot\a (+,0)}}-2i\t^{\a(-,0)}\p^A_{\a\dot\a}
-2i\bar\t_{\dot\a}^{(-,0)}\p^A_z, \q
\bar{D}^{(0,-)}_{\dot\a}=-\f{\p}{\p \bar\t^{\dot\a
(0,+)}}-2i\t^{\a(0,-)}\p^A_{\a\dot\a}-2i\bar\t_{\dot\a}^{(0,-)}\p^A_z~.$$

The harmonic derivatives $D^{(q_1,q_2)}$ in the analytic basis are
presented in Appendix (\ref{Dlong}). It is obvious that the property
(\ref{hshort}) in the $\tau$-frame is becoming a requirement of the
harmonic analyticity
$D^{(++,0)}W^{(+,+)}=D^{(\pm,\mp)}W^{(+,+)}=D^{(0,++)}W^{(+,+)}=0$
in  the $\lambda$-frame.

Similarly, we can consider the solution of the Grassmann constraints
(\ref{harmonic constraints1})-(\ref{harmonic constraints3}) for
other superfields $W^{(q_1,q_2)}$ by passing to the corresponding
analytic coordinates.

\subsection{Superfield action}
In this subsection we formulate the superfield action. We show that
such a superfield action is written in terms of superfield
$W^{(+,+)}$ and exactly reproduces the component action (\ref{act
comp}).

The action under consideration must be gauge invariant, ${\cal N}=4$
supersymmetric and invariant under central charge transformation. To
find such action, one uses a prescription, which was formulated in
the $\cN=2$ central charge harmonic superspace \cite{N2harmonic}. We
will see that this prescription perfectly works in $USp(4)$, $\cN=4$
harmonic superspace.

We propose the superfield action for $USp(4)$ SYM theory in the form
\be\label{actW} S\sim  {\tr}\int
d\zeta^{(-4,-4)}du((\t^{(+,0)})^2-(\bar\t^{(+,0)})^2)((\t^{(0,+)})^2-(\bar\t^{(0,+)})^2){\cal
L}^{(2,2)}~, \ee and show that it reproduces the component action
from \cite{SSW80}. Here \be\label{L^{2,2}} {\cal
L}^{(2,2)}=W^{(+,+)}W^{(+,+)}~, \ee is an analytic (\ref{harmonic
constraints}) and harmonically 'short' superfield (\ref{hshort}).
The analytic superspace dimensionless integration measure looks like
\be\label{anmeasure}d\zeta^{(-4,-4)}du=d^4x_Ad^2\t^{(+,0)}d^2\t^{(0,+)}d^2\bar\t^{(+,0)}d^2\bar\t^{(0,+)}du~,\ee
where $du$ denotes the left-right invariant measure on the
$USp(4)/U(1)\times U(1)$ coset\footnote{See the details in Appendix
B.}.

The action (\ref{actW}) is obviously gauge invariant. Also, this
action is ${\cN=4}$ supersymmetric. The proof of this statement is
analogous to one in $\cN=2$ theory with intrinsic central charges
\cite{N2harmonic}. The $\cN=4$ supersymmetry coordinate
transformations in the theory under consideration are
\be \d
x_{\a\dot\a}^A=-2i(\epsilon_\a^{ (-,0)}\bar\t^{(+,0)}_{\dot\a}+
\epsilon_\a^{ (0,-)}\bar\t^{(0,+)}_{\dot\a}+\t_\a^{
(+,0)}\bar\epsilon^{(-,0)}_{\dot\a} +\t_\a^{
(0,+)}\bar\epsilon^{(0,-)}_{\dot\a})~,\ee
$$\d z_A=2i(\epsilon^{\a (-,0)}\t_\a^{(+,0)}+\epsilon^{\a (0,-)}\t_\a^{(0,+)}-
\bar\epsilon_{\dot\a}^{ (-,0)}\bar\t^{(+,0)\dot\a}-
\bar\epsilon_{\dot\a}^{ (0,-)}\bar\t^{(0,+)\dot\a}), \q \d\t_{\hat{\a}}^{(+,0)}=\epsilon_{\hat{\a}}^{(+,0)}~.$$
Under these transformations the action (\ref{actW}) transforms as follows
$$\d S|_{\epsilon^{(-,0)}}=\int d\zeta^{(-4,-4)}du\{((\t^{(+,0)})^2-(\bar\t^{(+,0)})^2)
((\t^{(0,+)})^2-(\bar\t^{(0,+)})^2)\d z_{A}\f{\p}{\p z_{A}}{\cal L}^{(2,2)}$$
$$-2(\epsilon^{(+,0)}\t^{(+,0)}-\bar\epsilon^{(+,0)}\bar\t^{(+,0)})((\t^{(0,+)})^2-(\bar\t^{(0,+)})^2){\cal L}^{(2,2)}\}~.$$
After integrating by parts and using the identities
$$D^{(++,0)}\d z_{A}=2i(\epsilon^{\a
(+,0)}\t_\a^{(+,0)}-\bar\epsilon_{\dot\a}^{
(+,0)}\bar\t^{(+,0)\dot\a}), \q\q D^{(0,++)}\d z_{A}
=2i(\epsilon^{\a (0,+)}\t_\a^{(0,+)}-\bar\epsilon_{\dot\a}^{
(0,+)}\bar\t^{(0,+)\dot\a})~,$$
one gets
$$\d S|_{\epsilon^{(-,0)}}=-\int d\zeta^{(-4,-4)}du((\t^{(+,0)})^2-(\bar\t^{(+,0)})^2)((\t^{(0,+)})^2
-(\bar\t^{(0,+)})^2)\d z_{A}D^{(++,0)}{\cal L}^{(2,2)}=0~.$$
Analogously $\d S|_{\epsilon^{(0,-)}}=0$. As a result, the action
(\ref{actW}) is $\cN=4$ supersymmetric.

Though the action (\ref{actW}) does not contain integration over
$z$, it actually is $z$-independent due to the identities
$D^{(++,0)}{\cal L}^{(2,2)}=D^{(0,++)}{\cal L}^{(2,2)}=0$. Indeed,
the $z$ derivative of (\ref{actW}) is
$$\f{\p}{\p z} S \sim i\int d\zeta^{(-4,-4)}du \q ((\t^{(0,+)})^2-(\bar\t^{(0,+)})^2)(\p^{(++,0)}-
2i\t^{(+,0)}\s^m\bar\t^{(+,0)}\p_m){\cal L}^{(2,2)}=0.~$$ It is easy
to see that the integrand in above relation is a total $x$ and
$u$-derivative and disappears upon integration. Therefore this
action is invariant under central charge transformation $\d x_A^m=0,
\ \ \d z_A=\omega$ with  the rigid parameter $\omega$.

Now we rewrite the action (\ref{actW}) in component form. To do that we should integrate over
harmonic and over all anticommuting coordinates. We take into account that on a gauge
invariant quantities $D^2=\n^2$ \cite{1001}. Also one uses the integration rule
\be \int d\zeta^{(-4,-4)}du=\f{1}{256}\int d^4x
du^{(\pm,0)}du^{(0,\pm)}(\n^{(-,0)})^2(\bar{\n}^{(-,0)})^2(\n^{(0,-)})^2(\bar{\n}^{(0,-)})^2~.
\ee

First, one integrates over anticommuting coordinates $\t^{(0,+)},
\bar\t^{(0,+)}$  and gets\footnote{Here the
identities $\f14 (D^{(0,-})^2(\t^{(0,+)})^2=1$, $\f14
(\bar{D}^{(0,-)})^2(\bar\t^{(0,+)})^2=1$ and the definitions (\ref{harmBianch}), (\ref{harmBianch1}) are used.}
\be\label{chiral} S \sim
\tr\int d^4x du^{(\pm,0)}d^2\t^{(+,0)}{\cal L}^{(++,0)}
-\tr\int du^{(\pm,0)}d^4x d^2\bar\t^{(+,0)}{\cal L}^{(++,0)}~, \ee
where
\be\label{L2} {\cal L}^{(++,0)}=\frac{1}{4}\int du^{(0,\pm)}((\n^{(-,0)})^2 -
(\bar{\n}^{(-,0)})^2){\cal L}^{(2,2)} \ee
$$ =-\f12\int du^{(0,\pm)}(G^{(+,0)\a}G^{(+,0)}_{\a}+\bar{G}^{(+,0)\dot\a}\bar{G}^{(+,0)}_{\dot\a}-2iH^{(+,-)}W^{(+,+)})~.$$

Second, one integrates in (\ref{L2}) over
harmonics $u_i^{(0,\pm)}$ and finds
\be\label{L_{ij}}
{\cal
L}^{(++,0)}=u^{(+,0)(i}u^{(+,0)j)}{\cal L}_{ij}~,
\ee
where
$$
{\cal L}_{ij}=-\frac{1}{2}(G_i^{\a}G_{\a j}+\bar{G}_i^{\dot\a}\bar{G}_{\dot\a j}+\f{i}{2} H_{i}^{\ \ k}W_{jk})~.
$$

Third, using the table of integrals (\ref{int}) over harmonic
variables $u_i^{(\pm,0)}$ for the expression
$\f14(\n_{\hat{\a}}^{-,0})^2{\cal L}^{(++,0)}$ one gets
\be\label{act fin} S\sim -\f{1}{20}\tr\int d^4x (\n^{\a
(i}\n^{j)}_\a -\bar{\n}^{(i}_{\dot\a}\bar{\n}^{\dot\a j)}){\cal
L}_{ij}|_{\t=0}=\tr\int d^4x{\cal L}~, \ee where the integrand is as
follows:
\be\label{zoo} {\cal L}=-\f{1}{40}\{[\n^{\b i},\n_\b^j]
G^\a_i G_{\a j}+\n^{\b j}G^{\a i} \n_{\b i}G_{\a_j}-\n^{\b i}G^\a_i
\n_{\b }^jG_{\a_j} \ee
$$+[\n^{\b i},\n_\b^j] \bar{G}^{\dot\a}_i \bar{G}_{\dot\a j}+\n^{\b j}\bar{G}^{\dot\a i} \n_{\b i}\bar{G}_{\dot\a_j}-
\n^{\b i}\bar{G}^{\dot\a}_i \n_{\b }^j\bar{G}_{\dot\a_j}$$
$$+[\bar{\n}^{\dot\b i},\bar{\n}_{\dot\b}^j] G^\a_i G_{\a j}+\bar{\n}^{\dot\b j}G^{\a i} \bar{\n}_{\dot\b i}G_{\a_j}-
\bar{\n}^{\dot\b i}G^\a_i \bar{\n}_{\dot\b }^jG_{\a_j}$$
$$+[\bar{\n}^{\dot\b i},\bar{\n}_{\dot\b}^j] \bar{G}^{\dot\a}_i \bar{G}_{\dot\a j}+\bar{\n}^{\dot\b j}\bar{G}^{\dot\a i}
\bar{\n}_{\dot\b i}\bar{G}_{\dot\a_j}
-\bar{\n}^{\dot\b i}\bar{G}^{\dot\a}_i \bar{\n}_{\dot\b }^j\bar{G}_{\dot\a_j}$$
$$+\f{i}{2}[\n^{\a i},\n_\a^j]H_i^{\ \ l}W_{jl}+\f{i}{2}H_i^{\ \ l}[\n^{\a i},\n_\a^j]W_{jl}+
i\n^{\a}_jH_{il}\n_\a^iW^{jl}+i\n^{\a i}H_{i}^{\ \ l}\n_\a^jW_{jl}$$
$$+\f{i}{2}[\bar{\n}^{\dot\a i},\bar{\n}_{\dot\a}^j]H_i^{\ \ l}W_{jl}+\f{i}{2}H_i^{\ \ l}
[\bar{\n}^{\dot\a i},\bar{\n}_{\dot\a}^j]W_{jl}
+i\bar{\n}^{\dot\a}_jH_{il}\bar{\n}_{\dot\a}^iW^{jl}+i\bar{\n}^{\dot\a
i}H_{i}^{\ \ l}\bar{\n}_{\dot\a}^jW_{jl}\}~.$$
This expression
contains all the necessary terms corresponding to the component
action (\ref{act comp}). We can show, after some rather cumbersome calculations
using the identities (\ref{F-G}-\ref{Dir})\footnote{The results of calculations for each line in
(\ref{zoo}) are given in Appendix C.}, that the
action (\ref{act fin}) with a coefficient $\frac{1}{4}$ in the definition
(\ref{actW}) is rewritten in the form
\be\label{components}
{\cal L}=-\f14F^{mn}F_{mn}-\f12V^mV_m+\f18H^{ij}H_{ij}+\f18\n^m W_{ij}\n_m W^{ij}+
\f{1}{16}[W_{ik},W_j^{\ \ k}][W^i_{\ \ l},W^{jl}]
\ee
$$+iG^{\a i}\n_{\a\dot\a}\bar{G}^{\dot\a}_{ i}+
\f{i}{2}[W_{ik},{G}^{\a k}]G^i_\a+\f{i}{2}[W_{ik},\bar{G}_{\dot\a}^{
k}]\bar{G}^{\dot\a i}~.$$

It is easy to see that each term in the action (\ref{components})
has the corresponding analogous term in the action (\ref{act comp}),
that is eq. (\ref{components}) coincides with eq. (\ref{act comp})
up to the coefficients. But all such coefficients can be absorbed in
the redefinition of the  fields in (\ref{components}).  Besides, one
points out that the relation (\ref{DV}) at $z=0$ and switched off
anticommuting variables exactly reproduces the constraint
(\ref{costrV}). Therefore the action (\ref{components}) is
automatically accompanied by the constraint (\ref{costrV}). Thus,
all the transformation rules (\ref{supertrans}), (\ref{centrtrans})
and the constraint (\ref{costrV}) on the auxiliary fields $V_m,
H_{ij}$ are a consequence  of the Bianchi identities. As a result,
we finally derive the action (\ref{act comp}) from the superfield
action (\ref{actW}).

\section{Summary}

We have developed the harmonic superspace formulation of ${\cal
N}=4$ SYM theory with the rigid central charge. Component
formulation of this theory was given in \cite{SSW80}. We studied the
gauge theory in $USp(4)$, ${\cal N}=4$ superspace and showed that
all the constraints on the component field theory \cite{SSW80}, the
supersymmetry transformations and the central charge transformation
are the consequences of the Bianchi identities for the superfield
strengths. Also it was proved that the Lagrange multiple $A_5$ and
the expressions of auxiliary fields $V_m$ and $H_{ij}$ in terms of
$A_5$, used in \cite{SSW80}, have a natural origin as the conditions
of central charge independence of gauge superconnections
$\Gamma_{m,z}$ and superstrength $W_{ij}$. Under these conditions,
some of Bianchi identities are converted into equations of motion
for conventional ${\cal N}=4$ SYM theory and thus the $ USp(4)$ SYM
theory is transformed into conventional $SU(4)$ SYM theory.

We constructed the $USp(4)$, $\cN=4$ harmonic superspace and several
corresponding analytic subspaces. It is proved that Bianchi
identities in conventional and harmonic ${\cN=4}$ superspaces with
central charge provide the superspace treatment of all components of
the model \cite{SSW80}. The harmonic superspace under consideration
allows us to introduce the analytic superfield strength $W^{(+,+)}$
(see Subsection 3.1), which is a basic object for a analytic
superfield Lagrangian ${\cal L}^{(2,2)}$ (see Subsection 3.2). Gauge
invariant, $\cN =4$ supersymmetric action, invariant under the
central charge transformations is proposed and it is proven that
this action reproduces the component action given in \cite{SSW80}.

However, we should note that the superfield strength $W^{(+,+)}$ is
not expressed yet in terms of unconstrained superfield
prepotentials. The procedure, which allowed to construct harmonic
superspace formulation of $\cN=2$ and $\cN=3$ supersymmetric
theories in terms of the unconstrained analytic superfields, does
not work apparently, in its literal form, for $\cN=4$ SYM theory
with the central charge. Finding the prepotentials requires
developing the new approaches. Nevertheless we hope that the
approach developed in this paper will be useful for understanding
the possibilities to construct the unconstrained superfield
formulation $\cN=4$ SYM theory with central charge.

\section*{Acknowledgments }
The authors would like to thank E.A. Ivanov, I.B. Samsonov  and B.M.
Zupnik for discussions and N. Kawamoto, S.M. Kuzenko and K. Stelle
for comments. The work was partially supported by RFBR grant,
project No 12-02-00121 and by a grant for LRSS, project No
224.2012.2. The study was partially supported by the Ministry of
education and science of Russian Federation, project 14.B37.21.0774.
Also, I.L.B. acknowledges the partial support of the RFBR-Ukraine
grant, project No 11-02-90445, RFBR-DFG grant, project No
13-02-91330 and DFG grant, project No LE 838/12-1. N.G.P.\
acknowledges the partial support of the RFBR grant, project No
11-02-00242.

\section{Appendix A: Notations and Conventions}
\refstepcounter{section}
\def\theequation{A.\arabic{equation}}
\setcounter{equation}{0}
In this Appendix  we list the notations and conventions for spinor algebra  and some useful formulas
of the two-component spinor formalism \cite{GIOS}, \cite{Idea}.

The spinor indices are raised and lowered by means of
the antisymmetric symbols $\ve_{\a\b}, \ve_{\dot\a\dot\b}$ which have
the properties
\be\ve_{\a\g}\ve^{\g\b}=\d^\b_\a,\q
\ve_{\dot\a\dot\g}\ve^{\dot\g\dot\b}=\d^{\dot\b}_{\dot\a}~.\ee
The $USp(4)$ indices are raised and lowered by means of the
antisymmetric matrix $ \O_{ij},$ ($\O_{ik}\O^{kj}=\d^j_i$) according to the rule
\be  \psi^{\a i}=\ve^{\a\b}\O^{ij}\psi_{\b j}, \q \bar\psi^{\dot\a
i}=\ve^{\dot\a\dot\b}\O^{ij}\bar\psi_{\dot\b j}~. \ee
The matrix $\O_{ij}$ has the form
\be\label{O}\O_{ij}=\left(\begin{array}{cccc} 0 & 0 & 0 & 1 \\ 0 & 0 & 1 & 0 \\
0 & -1 & 0 & 0 \\  -1 & 0 & 0 & 0 \\\end{array} \right)~ .\ee

The products of anticommuting spinors are defined as
follows:
$$\t^\a\t^\b=-\f12\ve^{\a\b}\t^2,\q  \t_\a\t_\b=\f12\ve_{\a\b}\t^2,\q  \bar\t_{\dot\a}\bar\t_{\dot\b}=
-\f12\ve_{\dot\a\dot\b}\bar\t^2, \q
 \bar\t^{\dot\a}\bar\t^{\dot\b}=\f12\ve^{\dot\a\dot\b}\bar\t^2~,$$
$$\psi_\a\chi_\b=\f12\ve_{\a\b}\psi^\d\chi_\d+\f12\psi_{(\a}\chi_{\b)},\q \bar\psi_{\dot\a}\bar\chi_{\dot\b}=
-\f12\ve_{\dot\a\dot\b}\bar\psi_{\dot\d}\bar\chi{\dot\d}+\f12\bar\psi_{(\dot\a}\bar\chi_{\dot\b)}~.$$

The sigma matrices $\s^m_{\a\dot\a},$ $\bar\s_m^{\dot\a\a}=\ve^{\a\b}\ve^{\dot\a\dot\b}\s^m_{\b\dot\b},$ $
\s_a\bar\s_b=\eta_{ab}-i\s_{ab}$
have the basic properties
\be
\s^b_{\a\dot\a}\bar\s_a^{\dot\a\a}=2\d^b_a, \q
\s^{m}_{\a\dot\a}\bar\s^{\dot\b\b}_m=2\d_\a^\b\d_{\dot\a}^{\dot\b}~,
\ee
$$\s_a\bar\s_b\s_c=\eta_{ab}\s_c-\eta_{ac}\s_b+\eta_{bc}\s_a - i\ve_{abcd}\s^d,\q \bar\s_a\s_b\bar\s_c=
\eta_{ab}\bar\s_c-\eta_{ac}\bar\s_b+\eta_{bc}\bar\s_a
+ i\ve_{abcd}\bar\s^d~,$$
$$\f{i}{2}\ve_{abcd}\s^{cd}=\s_{ab},\q \f{i}{2}\ve_{abcd}\bar\s^{cd}=-\bar\s_{ab}~, $$
$$
\s^{ab}_{\a\b}\s_{ab}^{\g\d}=4(\d_{\a}^{\g}\d_{\b}^{\d}+\d_{\a}^{\d}\d_{\b}^{\g}),
\q \bar\s_{ab}^{\dot\a\dot\b}\bar\s^{ab}_{\dot\g\dot\d}=4
(\d^{\dot\a}_{\dot\g}\d^{\dot\b}_{\dot\d}+\d^{\dot\a}_{\dot\d}\d^{\dot\b}_{\dot\g})~,$$
$$(\s_{mn})_{\a\b}(\s_{ab})^{\b\a}=2(\eta_{[ma}\eta_{n]b}+ i\ve_{mnab}),\q
(\bar\s_{mn})_{\dot\a\dot\b}(\bar\s_{ab})^{\dot\b\dot\a}=2(\eta_{[ma}\eta_{n]b}-
i\ve_{mnab})~.$$

The relation between spinor and vector representations looks like
\be x^{\dot\a \a}=x^a\bar\s_a^{\dot\a\a},\q
x^a=\f12x^{\dot\a\a}\s^a_{\a\dot\a}~, \ee
$$ F^{\a\b}=(\s_{mn})^{\a\b}F^{mn},\q \bar{F}^{\dot\a\dot\b}=
(\bar\s_{mn})^{\dot\a\dot\b}F^{mn},\q F_{mn}=\f18 F^{\a\b}(\s_{mn})_{\a\b}+
\f18\bar{F}^{\dot\a\dot\b}(\bar\s_{mn})_{\dot\a\dot\b}~.$$

The rigid covariant derivatives have the form
\be D_m=\p_m,\q D_\a^i=\f{\p}{\p\t^\a_i}+i\bar\t^{\dot\alpha i}\p_{\a\dot\alpha}-i\t^i_\a\p_z, \q
\bar{D}_{\dot\a i}=-\f{\p}{\p\bar\t^{\dot\a
i}}-i\t^\alpha_i\p_{\a\dot\alpha}-i\bar\t_{\dot\a i}\p_{{z}}~,\ee
and satisfy the relations
\be\{D_\a^i,\bar{D}_{\dot\a j}\}=-2i\d^i_j\p_{\a\dot\a},\q \{D^i_\a,D^j_\b\}=-2i\ve_{\a\b}\O^{ij}\p_z, \q
\{\bar{D}_{\dot\a i},\bar{D}_{\dot\b
j}\}=2i\ve_{\dot\a\dot\b}\O_{ij}\p_{{z}}~.\ee
The expressions for the
$\cN=4$ supersymmetry generators are
\be Q_\a^i=i\f{\p}{\p\t^\a_i}+\bar\t^{\dot\a i}\p_{\a\dot\a}-\t^i_\a\p_z, \q \bar{Q}_{\dot\a i}=
-i\f{\p}{\p\bar\t^{\dot\a i}}-\t^{\a}_{ i}\p_{\a\dot\a}-\bar\t_{\dot\a i}\p_z~.
\ee


\section{Appendix B. Harmonics for $USp(4)/U(1)\times U(1)$}

\refstepcounter{section}
\def\theequation{B.\arabic{equation}}
\setcounter{equation}{0} Central charge breaks the $U(4)$ R-symmetry
group of the $\cN=4$ superalgebra down to $USp(4)$ \cite{S78}.
Therefore, to construct the corresponding harmonic superspace we
should define the $USp(4)$ harmonic variable. The various cosets of
the $USp(4)$ group were introduced and studied in
\cite{IKVO85}, \cite{BLS}. In the present work we are interested in
the harmonics on the $USp(4)/U(1)\times U(1)$ coset. In this
Appendix we describe the properties of such harmonics.

The $USp(4)$ harmonic variables are $4 \times 4$ unitary matrices
$u_i^I$ with unit determinant preserving the antisymmetric tensor (symplectic metric)
$\O^{ij} (\ref{O})$
\be  u_i^I\bar{u}_J^i=\d^I_J,\q
u^I_i\O^{ij}u^J_j=\O^{IJ},\ee
$$u^{Ii}=\O^{ik}u^I_k=\O^{IK}\bar{u}^i_K, \q u^I_i=\O_{ij}u^{Ij},\q \bar{u}^i_I=
\O_{IJ}u^J_k\O^{ki}~.
$$
It is convenient to label the harmonic variables by their U(1)
charges
$$u^1_i=u_i^{(+,0)},\q u^2_i=u_i^{(-,0)},\q u^3_i=u_i^{(0,+)},\q u^4_i=u_i^{(0,-)}~.
\footnote{$$\bar{u}^{i(+,0)}=\O^{ij}u_j^{(+,0)}, \q
\bar{u}^{i(-,0)}=\O^{ij}u_j^{(-,0)}, \q
\bar{u}^{i(0,+)}=\O^{ij}u_j^{(0,+)}, \q
\bar{u}^{i(0,-)}=\O^{ij}u_j^{(0,-)}.$$}$$ Then, the basic relations
for the harmonics can be written as orthogonality \be
\bar{u}^{i(+,0)}u_i^{(-,0)}=\bar{u}^{i(0,+)}u_i^{(0,-)}=1~,\ee
$$ \bar{u}^{i(0,+)}u_i^{(+,0)}=\bar{u}^{i(0,-)}u_i^{(+,0)}=\bar{u}^{i(-,0)}u_i^{(0,+)}=\bar{u}^{i(0,-)}u_i^{(-,0)}=0~,$$
and completeness conditions
\be
\det(u)=\ve^{ijkl}u_i^{(+,0)}u_j^{(-,0)}u_k^{(0,+)}u_l^{(0,-)}=1~,\ee
$$\bar{u}^{i(+,0)}u_j^{(-,0)}-\bar{u}^{i(-,0)}u_j^{(+,0)}+\bar{u}^{i(0,+)}u_j^{(0,-)}-\bar{u}^{i(0,-)}u_j^{(0,+)}=\d^i_j~.$$
These basic relations give us a possibility to convert
$USp(4)$ indices into $U(1)$ ones and vice versa, namely
\be\psi_i=u^{(+,0)}_i\psi^{(-,0)}-u^{(-,0)}_i\psi^{(+,0)}+u^{(0,+)}_i\psi^{(0,-)}-u^{(0,-)}_i\psi^{(0,+)}~.\ee
Thus, the harmonic variable may  be treated as the vielbeins which will be
used to transform $USp(4)$ tensors into $USp(4)$ singlets.

Besides the usual complex conjugation
$$\overline{(u_i^{(\pm,0)})}=\mp u^{(\mp,0)i}, \q \overline{(u_i^{(0,\pm)})}=\mp u^{(0,\mp)i},$$
there is the following 'tilde'-conjugation for harmonics
\cite{IKVO85}, \cite{BLS}
\be\label{combconj}\widetilde{u_i^{(\pm,0)}}=u^{(0,\pm)i},\q
\widetilde{u^{(\pm,0)i}}=-u_i^{(0,\pm)}, \q
\widetilde{u_i^{(0,\pm)}}=u^{(\pm,0)i},\q
\widetilde{u^{(0,\pm)i}}=-u_i^{(\pm,0)}~.\ee This operation involves
a complex conjugate and one of the reflections in the Weyl algebra
$usp(4).$ It acts only on the tangent indices of harmonics and can
be selected in several ways. Among the many possible definitions of
combined conjugation, (\ref{combconj}) is most suitable for our
purposes. It is the conjugation  which allows us to define real
objects in harmonic superspace with $USp(4)$ harmonics.

There are also important identities with harmonics
\be
\bar{u}^{i(+,0)}\bar{u}^{j(-,0)}\ve_{ijkl}=+
u^{(0,+)}_{[k}u^{(0,-)}_{l]}, \q
\bar{u}^{i(0,+)}\bar{u}^{j(0,-)}\ve_{ijkl}=u^{(+,0)}_{[k}u^{(-,0)}_{l]}~,
\ee
$$\bar{u}^{i(+,0)}\bar{u}^{j(0,-)}\ve_{ijkl}=-u^{(+,0)}_{[k}u^{(0,-)}_{l]},
\q \bar{u}^{i(0,+)}\bar{u}^{j(-,0)}\ve_{ijkl}=-u^{(0,+)}_{[k}u^{(-,0)}_{l]}~, $$
$$\bar{u}^{i(+,0)}\bar{u}^{j(0,+)}\ve_{ijkl}=-u^{(+,0)}_{[k}u^{(0,+)}_{l]},
\q\bar{u}^{i(-,0)}\bar{u}^{j(0,-)}\ve_{ijkl}=-u^{(-,0)}_{[k}u^{(0,-)}_{l]}~. $$

For the anticommuting variables we define the harmonic projections
\be\t^I_\a=-\bar{u}^{Ii}\t_{\a i}=u^I_i\t^i_\a, \q
\bar\t^I_{\dot\a}=u^I_i\bar\t^i_{\dot\a}=-\bar{u}^{I
i}\bar\t_{\dot\a i}~,\ee
and following the rules of conjugation
\be\widetilde{\t_\a^{(\pm,0)}}=\bar\t_{\dot\a}^{(0,\pm)}, \q
\widetilde{\t_\a^{(0,\pm)}}=\bar\t_{\dot\a}^{(\pm,0)}, \q
\widetilde{\bar\t_{\dot\a}^{0,\pm}}=-\t_\a^{(\pm,0)},\q
\widetilde{\bar\t_{\dot\a}^{\pm,0}}=-\t_\a^{(0,\pm)}~.\ee
Analogously, we define the harmonic projections for covariant spinor derivatives
\be
D^{(\pm,0)}_\a=\pm\f{\p}{\p
\t^{(\mp,0)\a}}+i\bar\t^{(\pm,0)\dot\a}\p_{\a\dot\a}-i\t^{(\pm,0)}_\a\p_z,\q
D^{(0,\pm)}_\a=\pm\f{\p}{\p
\t^{(0,\mp)\a}}+i\bar\t^{(0,\pm)\dot\a}\p_{\a\dot\a}-i\t^{(0,\pm)}_\a\p_z~,\ee
$$\bar{D}^{\pm,0}_{\dot\a}=\pm\f{\p}{\p \bar\t^{\dot\a (\mp,0)}}-i\t^{\a(\pm,0)}\p_{\a\dot\a}
-i\bar\t_{\dot\a}^{(\pm,0)}\p_z, \q
\bar{D}^{0,\pm}_{\dot\a}=\pm\f{\p}{\p \bar\t^{\dot\a
(0,\mp)}}-i\t^{\a(0,\pm)}\p_{\a\dot\a}-i\bar\t_{\dot\a}^{(0,\pm)}\p_z~.$$
They are also related by the conjugation as follows
$$\widetilde{D^{(\pm,0)}_\a}=-\bar{D}_{\dot\a}^{(0,\pm)}, \q \widetilde{D^{(0,\pm)}_\a}=-\bar{D}_{\dot\a}^{(\pm,0)},\q
\widetilde{\bar{D}^{(\pm,0)}_{\dot\a}}={D}_{\a}^{(0,\pm)},\q
\widetilde{\bar{D}^{(0,\pm)}_{\dot\a}}={D}_{\a}^{(\pm,0)}~.$$

The right action of $usp(4)$ algebra on  a space of harmonics $'u'$ is generated
by the  differential operators (\ref{Hpart}), (\ref{J}).
The commutation relations among these operators show that
the $\p^{(++,0)}, \p^{(-,+)}, \p^{(+,+)}, \p^{(0,++)}$ are rising
operators, $\p^{(--,0)}, \p^{(+,-)}, \p^{(-,-)}, \p^{(0,--)}$ are
lowering ones and $j_1,j_2$ are the Cartan generators in $usp(4)$
algebra. There are also the following reality properties
\be\widetilde{\p^{(++,0)}}=\p^{(0,++)},\q
\widetilde{\p^{(--,0)}}=\p^{(0,--)}, \q
\widetilde{\p^{(+,+)}}=\p^{(+,+)}, \q
\widetilde{\p^{(+,-)}}=-\p^{(-,+)}!,\ee
$$ \widetilde{\p^{(0,++)}}=\p^{(++,0)},\q \widetilde{\p^{(-,+)}}=-\p^{(+,-)}~.$$

It is easy to check that these derivatives satisfy
the relations
\be\label{harmalg}[j_1,
\p^{(q_1,q_2)}]=q_1\p^{(q_1,q_2)}, \q\q [j_2,
\p^{(q_1,q_2)}]=q_2\p^{(q_1,q_2)}~,\ee
$$[\p^{(++,0)}, \p^{(--,0)}]=j_1, \q [\p^{(0,++)}, \p^{(0,--)}]=j_2~,$$
$$[\p^{(+,+)}, \p^{(-,-)}]=j_1+j_2, \q [\p^{(+,-)}, \p^{(-,+)}]=-j_1+j_2~,$$
$$[\p^{(++,0)}, \p^{(-,+)}]=\p^{(+,+)}, \q [\p^{(+,+)}, \p^{(-,+)}]=2\p^{(0,++)}~, $$
$$[\p^{(+,-)}, \p^{(--,0)}]=-\p^{(-,-)},  \q [\p^{(+,-)}, \p^{(-,-)}]=-2\p^{(0, --)}~,$$
$$\q [\p^{(+,-)}, \p^{(0,++)}]=\p^{(+,+)},\q [\p^{(+,-)}, \p^{(+,+)}]=
2\p^{(++,0)}, \q [\p^{(+,+)}, \p^{(--,0)}]=-\p^{(-,+)}~, $$
$$ [\p^{(0,++)}, \p^{(-,-)}]=-\p^{(-,+)}~,$$
$$ [\p^{(-,+)}, \p^{(-,-)}]=2\p^{(--,0)},
\q [\p^{(+,+)}, \p^{(0,--)}]=\p^{(+,-)}, \q [\p^{(++,0)},
\p^{(-,-)}]=\p^{(+,-)}~,$$
$$ \q [\p^{(-,+)}, \p^{(0,--)}]=\p^{(-,-)}~,$$
where
\be\label{J} j_1=u^{(+,0)}_i\f{\p}{\p u^{(+,0)}_i}-u^{(-,0)}_i\f{\p}{\p
u^{(-,0)}_i},\q j_2=u^{(0,+)}_i\f{\p}{\p
u^{(0,+)}_i}-u^{(0,-)}_i\f{\p}{\p u^{(0,-)}_i}~.\ee

It is also useful to find the commutation relations of  $R$-
symmetry generators with harmonic projections of the spinor covariant
derivatives
\be\label{flatD}
[\p^{(\pm\pm,0)},D_{\hat{\a}}^{(\mp,0)}]=D_{\hat{\a}}^{(\pm,0)},\q
[\p^{(0,\pm\pm)},D_{\hat{\a}}^{(0,\mp)}]=D_{\hat{\a}}^{(0,\pm)}~,\ee
$$[\p^{(\pm,\pm)},D_{\hat{\a}}^{(0,\mp)}]=D_{\hat{\a}}^{(\pm,0)},\q [\p^{(\pm,\pm)},D_{\hat{\a}}^{(\mp,0)}]
=D_{\hat{\a}}^{(0,\pm)}~,$$
$$[\p^{(+,-)},D_{\hat{\a}}^{(0,+)}]=D_{\hat{\a}}^{(+,0)},\q [\p^{(+,-)},D_{\hat{\a}}^{(-,0)}]
=-D_{\hat{\a}}^{(0,-)}~,$$
$$[\p^{(-,+)},D_{\hat{\a}}^{(+,0)}]=-D_{\hat{\a}}^{(0,+)},\q [\p^{(-,+)},D_{\hat{\a}}^{(0,-)}]
=D_{\hat{\a}}^{(-,0)}~.$$
Besides, all derivatives are eigenvectors of the charge operators
$j_1,j_2.$

In the analytic basis (\ref{anbasis}) the operators of the algebra
$usp(4)$ read as:
\be\label{Dlong}
D^{(++,0)}=\p^{(++,0)}+\t^{(+,0)\a}\f{\p}{\p\t^{(-,0)\a}}+
\bar\t^{(+,0)\dot\a}\f{\p}{\p\bar\t^{(-,0)\dot\a}}-2i\t^{(+,0)}\not\!\p\bar\t^{(+,0)}
+i(\t^{(+,0)\a}\t^{(+,0)}_{\a}-\bar\t^{(+,0)}_{\dot\a}\bar\t^{(+,0)\dot\a})\p_z~,\ee
$$D^{(0,++)}=\p^{(0,++)}+\t^{(0,+)\a}\f{\p}{\p\t^{(0,-)\a}}+\bar\t^{(0,+)\dot\a}\f{\p}{\p\bar\t^{(0,-)\dot\a}}
-2i\t^{(0,+)}\not\!\p\bar\t^{(0,+)}
+i(\t^{(0,+)\a}\t^{(0,+)}_{\a}-\bar\t^{(0,+)}_{\dot\a}\bar\t^{(0,+)\dot\a})\p_z~,$$
$$D^{(\pm,\pm)}=\p^{\pm,\pm)}+\t^{(\pm,0)\a}\f{\p}{\p\t^{(0,\mp)\a}}+\t^{(0,\pm)\a}\f{\p}{\p\t^{(\mp,0)\a}}
+\bar\t^{(\pm,0)\dot\a}\f{\p}{\p\bar\t^{(0,\mp)\dot\a}}
+\bar\t^{(0,\pm)\dot\a}\f{\p}{\p\bar\t^{(\mp,0)\dot\a}}~,$$
$$-2i\t^{(\pm,0)}\not\!\p\bar\t^{(0,\pm)}-2i\t^{(0,\pm)}\not\!\p\bar\t^{(\pm,0)}
+2i(\t^{(\pm,0)\a}\t^{(0,\pm)}_{\a}-\bar\t^{(0,\pm)}_{\dot\a}\bar\t^{(\pm,0)\dot\a})\p_z~,$$
$$D^{(\pm,\mp)}=\p^{(\pm,\mp)}+\t^{(\pm,0)}\f{\p}{\p\t^{(0,\pm)}}+\bar\t^{(\pm,0)}\f{\p}{\p\bar\t^{(0,\pm)}}
-\t^{(0,\mp)}\f{\p}{\p\t^{(\mp,0)}}
-\bar\t^{(0,\mp)}\f{\p}{\p\bar\t^{(\mp,0)}}~,$$
$$D^{(--,0)}=\p^{(--,0)}+\t^{(-,0)}\f{\p}{\p\t^{(+,0)}}+\bar\t^{(-,0)}\f{\p}{\p\bar\t^{(+,0)}}-2i\t^{(-,0)}\not\!\p\bar\t^{(-,0)}
+i(\t^{(-,0)\a}\t_\a^{(-,0)}-\bar\t^{(-,0)}_{\dot\a}\bar\t^{(-,0)\dot\a})\p_z~,$$
$$D^{(0,--)}=\p^{(0,--)}+\t^{(0,-)}\f{\p}{\p\t^{(0,+)}}+\bar\t^{(0,-)}\f{\p}{\p\bar\t^{(0,+)}}
-2i\t^{(0,-)}\not\!\p\bar\t^{(0,-)}+i(\t^{(0,-)}\t^{(0,-)}-
\bar\t^{(0,-)}\bar\t^{(0,-)})\p_z~,$$
$$J_1=j_1+\t^{(+,0)}\f{\p}{\p\t^{(+,0)}}+\bar\t^{(+,0)}\f{\p}{\p\bar\t^{(+,0)}}-\t^{(-,0)}\f{\p}{\p\t^{(-,0)}}-
\bar\t^{(-,0)}\f{\p}{\p\bar\t^{(-,0)}}~,$$
$$J_2=j_2+\t^{(0,+)}\f{\p}{\p\t^{(0,+)}}+\bar\t^{(0,+)}\f{\p}{\p\bar\t^{(0,+)}}-\t^{(0,-)}\f{\p}{\p\t^{(0,-)}}-
\bar\t^{(0,-)}\f{\p}{\p\bar\t^{(0,-)}}~.$$

The Grassmann and harmonic measure of integration over the $USp(4)$
analytic harmonic superspace is $d\zeta^{(-4,-4)}= d^4x_A
d^2\t^{(+,0)}d^2\bar\t^{(+,0)}d^2\t^{(0,+)}d^2\bar\t^{(0,+)}du$,
where the superscript $(-4,-4)$ refers to the $(q_1,q_2)$ charges.
The measure is normalized so that
\be\label{measure}
\int d\zeta^{(-4,-4)}(\t^{(+,0)})^2(\t^{(0,+)})^2(\bar\t^{(+,0)})^2(\bar\t^{(0,+)})^2=1~.
\ee
Functions of harmonics are defined by their harmonic expansion
\be
f^{(q_1,q_2)}(u)=\sum_{k_1,k_2=0}^{\infty}f^{(i_1\ldots i_{2k_1+q_1})(j_1\ldots
j_{2k_2+q_2})}u^{(+,0)}_{(i_1}\ldots u^{(+,0)}_{i_{k_1}+q_1}u^{(-,0)}_{i_{k_1+q_1+1}}
\ldots u^{(-,0)}_{i_{2k_1+q_1})}\ee
$$\times u^{(0,+)}_{(j_1}\ldots
u^{(0,+)}_{j_{k_2}+q_2}u^{(0,-)}_{j_{k_2+q_2+1}} \ldots
u^{(0,-)}_{j_{2k_2+q_2})}~.
$$
The direct consequences of the above definitions are the following
rules
$$
\p^{(++,0)}f^{(q_1,q_2)}(u)=0 \rightarrow f^{(q_1,q_2)}(u)=0,
q_1<0;\q \p^{(0,++)}f^{(q_1,q_2)}(u)=0 \rightarrow
f^{(q_1,q_2)}(u)=0, q_2<0~.$$
The harmonic integral is defined to
select the $USp(4)$ singlet
\be
\int du 1=1, \q \int du f^{(q_1,q_2)}(u)=\d^{(q_1,0)}\d^{(0,q_2)}f^{(q_1,q_2)}(0)~.
\ee
Using this definition and the reduction identities
\cite{GIOS} one can derive the integral table
\be\label{int}
\int du u^{(+,0)i}u^{(-,0)}_j=\f14\d^i_j \q \int du u^{(+,0)i}u^{(+,0)j}
u^{(-,0)}_{k} u^{(-,0)}_l=\f{1}{4\cdot 5}\d^{ij}_{(kl)},
\q\mbox{etc.}
\ee

\section{Appendix C. }

\refstepcounter{section}
\def\theequation{C.\arabic{equation}}
\setcounter{equation}{0}

The results of the transformations of each line in (\ref{zoo}) are
expressed as follows

$\bullet$
$$20i\n_{\dot\a\a}\bar{G}^{\dot\a}_{ i}G^{i\a}-2H^{ij}H_{ij}+5F^{\a\b}F_{\a\b}-[W_{ik},W_j^{\ \ k}][W^i_{\ \ l},W^{jl}]~,$$

$\bullet$
$$24i[W_{ik},\bar{G}^{\dot\a k}]\bar{G}^i_{\dot\a}-20i\n_{\dot\a\a}G^\a{ i}\bar{G}^{i\dot\a}+
20V^{\dot\a\a}V_{\a\dot\a}-\n^{\dot\a\a}W^{ij}\n_{\a\dot\a}W_{ij}~,$$

$\bullet$
$$-24i[W_{ik},{G}^{\a k}]{G}^i_{\a}+20i\n_{\a\dot\a}\bar{G}^{\dot\a}_{ i}{G}^{\a i}+20V^{\dot\a\a}V_{\a\dot\a}-
\n^{\dot\a\a}W^{ij}\n_{\a\dot\a}W_{ij}~,$$

$\bullet$
$$20i\n_{\a\dot\a}{G}^{\a i}
\bar{G}^{\dot\a}_i-2H^{ij}H_{ij}+5F^{\dot\a\dot\b}F_{\dot\a\dot\b}-[W_{ik},W_j^{\ \ k}][W^i_{\ \ l},W^{jl}]~,$$

$\bullet$
$$40i\n_{\a\dot\a}\bar{G}^{\dot\a}_iG^{\a i}-8H^{ij}H_{ij}-4[W_{ik},W_j^{\ k}][W^i_{\ l},W^{jl}]-8\n^mW_{ij}\n_mW^{ij}$$
$$-16i[W_{ik},\bar{G}^k_{\dot\a}]\bar{G}^{\dot\a i}+40i[W_{ik},G^k_\a]G^{\a i}~,$$

$\bullet$
$$-40i\n_{\a\dot\a}G^\a_i\bar{G}^{\dot\a i}-8H^{ij}H_{ij}-4[W_{ik},W_j^{\ k}][W^i_{\ l},W^{jl}]-8\n^mW_{ij}\n_mW^{ij}$$
$$-16i[W_{ik},{G}^{\a k}]G^i_\a-40i[W_{ik},\bar{G}^k_{\dot\a}]\bar{G}^{\dot\a i}~.$$

\bigskip

\end{document}